\documentclass[11pt,showpacs]{revtex4}
\input epsf.sty

\usepackage[sort&compress]{natbib}
\usepackage{color}

\def\rr#1{\textcolor{red}{#1}}

\def\mn#1{\marginpar[\tiny{\rr{#1}}]{\tiny{\rr{#1}}}}

\def\comment#1{}

\def\mn#1{*\marginpar{*\tiny{#1}}}
\def\mn#1{}
\def\E{{\mathcal E}}

\usepackage{graphicx}
\begin{document}

\title{Critical fermion density for restoring spontaneously broken symmetry}
\author{Hagen Kleinert$^{(a,b)}$ and She-Sheng Xue$^{(b,c)}$}
\affiliation{$^{(a)}$Institut f{\"u}r Theoretische Physik, Freie Universit\"at Berlin, 14195 Berlin, Germany}
\affiliation{$^{(b)}$ICRANeT Piazzale della Repubblica, 10 -65122, Pescara, Italy}
\affiliation{$^{(c)}$
Dipartimento di Fisica, University of Rome ``La Sapienza", P.le A. Moro 5, 00185 Rome, Italy}

%\email{kleinert@physik.fu-berlin.de  ruffini@icra.it  xue@icra.it}

\date{Received version \today}

\begin{abstract}
We show how the phenomenon of spontaneous symmetry breakdown
is affected by the presence of a sea of fermions in the system. When its
density exceeds a critical value,
the broken symmetry can be restored. We calculate the critical value
and discuss the consequences for three different physical systems:
First, for the standard model of particle physics, where
the spontaneous symmetry breakdown leads nonzero masses of intermediate gauge 
bosons and fermions. The symmetry restoration 
will greatly enhance various processes with
dramatic consequences for the early universe. 
Second, for the Gell-Mann--L\`evy $\sigma$-model of nuclear physics,
where the symmetry breakdown gives rise
to the nucleon and meson masses. The symmetry restoration 
may have important consequences for formation or collapse of stellar cores. 
Third, for the superconductive phase of condensed-matter, where
the BCS condensate at low-temperature 
may be destroyed by a too large electron density. 
\end{abstract}

\pacs{11.30.Qc, 11.15.Ex,98.80.Cq,97.60.Lf}%,74.20.Fg

\maketitle

%\vskip0.1cm
\noindent
{\bf Introduction.}
\hskip0.1cm
Spontaneous symmetry breakdown 
is one of most important 
phenomena in many areas of modern physics \cite{nj}. In metals, it affects 
particle number conservation and gives rise to an energy gap that is 
responsible for superconductivity and the Meissner effect \cite{TINKHAMBOOK}. 
In the standard model (SM) of particle physics, spontaneous symmetry 
breakdown  produces  the  masses of elementary 
fermions and vector bosons, 
the latter being responsible for the 
weakness of weak interactions  \cite{njl}. Recall that electromagnetic 
cross sections are of the order of $\alpha^2/E^2$, 
where $E$ is the energy of the scattering particles, 
and $\alpha=e ^2/\hbar c~\approx 1/137$ the fine structure constant.
In weak interactions, these are reduced to  $\alpha^2/M_W^2$ or $\alpha^2/M_Z^2$
where vector boson masses $M_W\approx 80.4\,$GeV and $M_Z\approx 91.19\,$GeV.
This causes the slow
reaction dynamics of many energy-production processes 
in astrophysics.
In the sun, in which $E$ is of the order of 
MeV, it is thanks to the largeness of $M_{W,Z}$ that
allows a star to live for
several billion years. If $M_{W,Z}$ would drop to zero in a
phase transition  of starts,
suns could disappear in short supernova-like explosions.

Recall that in the Gell-Mann-L\`evy $\sigma$-model of nuclear physics, a
 spontaneously broken chiral  symmetry is responsible for the nucleon and meson masses, with the pion appearing 
as a would-be Nambu-Goldstone boson. The breakdown
explains the low pion mass, 
the observed magnitude of the partial conservation of the axial current (PCAC), 
and the associated pion decay constant.

Another phenomenon where a spontaneously broken symmetry and generated mass
has dramatic consequences is superconductivity.
There the mass manifests itself as an energy gap of the conduction 
electrons, that is responsible for the infinite conductivity below a certain critical temperature $T_c$.

In a Fermi liquid, a spontaneously broken symmetry is generally due to an attractive %(negative) 
potential of a system that causes a dynamical (Cooper) instability, 
which
 produces a low-mass pair bound state.
 On the other hand, the temperature, which represents 
 positive thermal energy and pressure of the system, plays the role of 
restoring spontaneously broken symmetries. As a result, there exists
a 
critical temperature $T_c$ with a second-order phase transition, below which 
there exist a
symmetric phase. Temperature or pressure can destroy this.
Increasing the fermion density
 causes a higher %{\it degenerate} 
pressure of the system, 
which plays the same role as increasing the temperature 
and restores 
the spontaneously broken 
symmetries. If the fermion density is larger than a critical value 
$n_c$, the nonzero condensation returns to zero and the symmetric 
situation is recovered. 
~\\
In this Letter, we shall demonstrate how this
phenomenon occurs and calculate the critical values of fermion density 
%$n_c$ in above 
where it happens.
We shall 
 discuss the various physical consequences. In addition, we point out the completely opposite directions of 
 influence of higher particle density 
for fermions and bosons with   
spontaneously broken symmetries.

%\vskip0.1cm
\noindent
{\bf Symmetry restoration in
U(1)-Higgs Model.}
\hskip0.1cm
To illustrate the basic  mechanism,
consider first the simplest model with a complex scalar field $\phi$ of the $U(1)$-symmetric field potential 
\begin{eqnarray}
V(\phi^*\phi)=\mu^2 \phi^*\phi +\lambda (\phi^*\phi)^2,~~~~\mu^2<0, ~~~~\lambda >0\,.
\label{hp}
\end{eqnarray}
We assume  that
$\mu^2<0$, so that the field $\phi$ has
a nonzero expectation value $\langle \phi\rangle\equiv \langle |\phi|\rangle$:
\begin{eqnarray}
\langle \phi\rangle= \sqrt{-\mu^2/2\lambda} \neq 0\,.
\label{eve}
\end{eqnarray} 
In the ground state, the U(1)-symmetry
is broken.
The complex scalar field is usually charged, 
and coupled to a gauge field, with an 
Euclidean action
\begin{eqnarray}
{\cal A}=\int d^4x\left\{ \frac{1}{2}
|D_\mu \phi|^2+V(\phi^*\phi)+\frac{1}{4}F_{\mu \nu }^2\right\} \,,
\nonumber
%\label{@}
\end{eqnarray}
where $F_{\mu \nu}=
\partial _\mu A_ \nu
-\partial _\nu A_ \mu $ is the
electromagnetic field tensor, %of the $A^\mu$-field
and
$D_\mu\equiv  \partial_\mu-ieA_\mu$
denotes the
the gauge-covariant derivatives.
The $A^\mu$-boson has a Meissner-Higgs mass
$m_A^2=e^2\langle \phi\rangle^2$.

Relativistic fermions are coupled by an action
\begin{eqnarray}
{\mathcal A}_{\rm f}&=&\int d^4x \bar\psi [\gamma_\mu (i\partial^\mu-e A^\mu) - m(\phi)]\psi\,,
\label{l}
\end{eqnarray}
where 
\begin{eqnarray}
m(\phi)=m_0+  g\phi\,,
\label{mphi}
\end{eqnarray}
$m_0$ are the bare fermion masses, and $g$ is a Yukawa coupling. 
In the ground state with $\langle \phi\rangle\neq 0$, the observed fermion masses are given by $m(\langle \phi\rangle)=m_0+  g\langle \phi\rangle$. 
The Dirac equation for the fermion field $\psi$ leads to the Pauli equation
\begin{eqnarray}
\Big[(i\partial^\mu-e A^\mu)^2 -\frac{e}{2}\sigma_{\mu\nu}F^{\mu\nu}- m^2(\phi)\Big]\psi=0,
\label{de}
\end{eqnarray}
where $e\sigma_{\mu\nu}F^{\mu\nu}=2e(i{\bf \alpha}\cdot {\bf E} +{\bf \sigma}\cdot {\bf B})= 2e{\bf \sigma}\cdot {\bf B}$ for vanishing electric field ${\bf E}=0$. In the semi-classical description of external fields $\phi$ and constant ${\bf B}$ pointing in the $\hat z$-direction, the energy levels are given by  
\begin{eqnarray}
\omega^2({\bf p})
= m(\phi)^2 + p_z^2 + p_\perp^2\,,
\label{om}
\end{eqnarray}
where $p_\perp$ is the squared transverse momentum satisfying 
$p_\perp^2=2eB(n+1/2-\hat s)$, $n=0,1,2,\cdots ,$ where $\hat s=\pm 1/2$ for a spin-$1/2$ fermion and $\hat s=0$ for a spin-$0$ boson. 

In thermal equilibrium at an inverse temperature
$ \beta $, fermions are described by a
Euclidean
action
\begin{eqnarray}
A&=&  \beta(H-\mu N)\,,
\label{l1}
\end{eqnarray}
where $\mu$ is the chemical potential.
%which at zero temperature determines the Fermi energy ${\mathcal E}_F=(P_F^2+m^2(\phi))^{1/2}+eA^0$.
%The value of $\mu$ is determined by 
The total particle-number and Hamiltonian are given by the phase space integrals
\begin{eqnarray}
N&=&
V\int \frac{d^3{\bf p}}{(2\pi)^3}\hat N_{{\bf p}}=V
\int \frac{d^3{\bf p}}{(2\pi)^3}
a^\dagger({\bf p})a({\bf p})\label{n}\,,\\
H&=&
V\int
%_{|{\bf p}|\le p_F} 
\frac{d^3{\bf p}}{(2\pi)^3}\hat H_{{\bf p}}=
V\int
%_{|{\bf p}|\le p_F} 
\frac{d^3{\bf p}}{(2\pi)^3}
\omega({\bf p})a^\dagger({\bf p})a({\bf p})\,,
\label{h}
\end{eqnarray}
where $
a^\dagger({\bf p})$ and $a({\bf p})$
are the creation and annihilation operators
of particles in momentum-${\bf p}$. We have suppressed the spin index 
$\hat s$ and the sums over $\hat s$, for brevity.
In an externalmagnetic field, the momentum integrals 
 are 
quantized as follows: 
$\int d^2{\bf p}_\perp/(2\pi)^2=eB/(2\pi)^2\sum_{n,\hat s}$.
For fermions, there are two possible
occupation numbers of each state, so that the particle number operator 
of each momentum state $\hat N_{\bf p}=a^\dagger({\bf p})a({\bf p})$  
has eigenvalues zero or one. For bosons, on the other hand,
the operator $N_{\bf p}=a^\dagger({\bf p})a({\bf p})
$ comprises all integer values $0,1,2,\dots$.
Their energies
$\hat H_{\bf p}= \omega ({\bf p})\hat N_{\bf p}$ and
partition functions
for each momentum state are
\begin{eqnarray}
Z_{\bf p}^F
&=&\sum_{N_{\bf p}=0,1} e^{-\beta(H_{\bf p}-\mu N_{\bf p})}=1+e^{-\beta(\omega({\bf p})-\mu)}\,,
\label{f}\\
Z_{\bf p}^B &=&\sum_{N_{\bf p}} e^{-\beta(H_{\bf p}-\mu N_{\bf p})}=\left[1-e^{-\beta(\omega({\bf p})-\mu)}\right]^{-1}.\,
\label{b}
\end{eqnarray}
These yield the Fermi $(+)$ and Bose-Einstein $(-)$ distributions 
\begin{eqnarray}
\langle N_{\bf p} \rangle \!=\!\frac{1}{Z_{\bf p}^{F,B}}\sum_{N_{\bf p}} N_{\bf p} e^{-\beta(H_{\bf p}-\mu N_{\bf p})}
%=\frac{\partial \ln Z_{\bf p}^F}{\partial ( \beta \mu)}
=\frac{1}{e^{\beta(\omega({\bf p})-\mu)}\pm 1}\,.
\label{fd}
%\\
%\langle N_{\bf p} \rangle=\frac{1}{Z_{\bf p}^B}\sum_{N_{\bf p}}
%N_{\bf p} e^{-\beta(H_{\bf p}-\mu N_{\bf p})}
%=\frac{\partial \ln Z_{\bf p}^B}{\partial  ( \beta \mu)}=[e^{\beta(\omega({\bf p})-\mu)}-1]^{-1},~~{\rm boson}\,. \label{bd}
\end{eqnarray}
Each partition function
$Z_{\bf p}^{F,B}$
defines a grand-canonical free energy
$ \beta  \Omega _{\bf p}^{F,B}\equiv -\log Z_{\bf p}^{F,B}$.
The total grand-canonical free energy $\Omega$ is the sum of all possible microscopic states, determining the
pressure of the system $p= -\Omega /V$
\begin{eqnarray}
p^F &=& \frac{2}{\beta}\int \frac{d^3{\bf p}}{(2\pi)^3} \ln \left[1+e^{-\beta(\omega({\bf p})-\mu)}\right]\,, 
\label{fp}\\
p^B &=& -\frac{1}{\beta}\int \frac{d^3{\bf p}}{(2\pi)^3} \ln \left[1-e^{-\beta(\omega({\bf p})-\mu)}\right]\,,
\label{bp}
\end{eqnarray}
where the factor $2$ of Eq.~(\ref{fp}) accounts for the two spin directions of fermions.

We consider the case of fermions, the pressure terms depend via $m(\phi)$ of Eq.~(\ref{mphi}) on the scalar field $\phi$
and thus modify the field potential
(\ref{hp}), which in the presence of
fermions is changed  to
\begin{eqnarray}
V_{\rm eff} &=&
V(\phi^*\phi) -p\nonumber\\
&=&
V(\phi^*\phi)
\!-\!\frac{2}{\beta}\int \frac{d^3{\bf p}}{(2\pi)^3} \ln \left[1\!+\!e^{-\beta(\omega({\bf p})-\mu)}\right].
\label{effp}
\end{eqnarray}
Because the second term vanishes for large fields $\phi$, this effective potential is positive definite ($V_{\rm eff}>0$). However, the expectation value $\langle \phi \rangle$, i.e., the location of the ground state in the field configurations, can be changed depending on the chemical potential, temperature and magnetic field. In particular, the spontaneously broken
 $U(1)$-symmetry can be restored, if $\langle \phi \rangle$ 
becomes zero for some critical values $\mu_c$, $\beta_c$ and $B_c$.  In the following, we shall focus our attention
upon this restoration process and its observable consequences for  our Universe.

To 
start with we  assume 
 the temperature to be zero and
that there is no  magnetic field.
Then the effective potential (\ref{effp}) becomes
\begin{eqnarray}\!\!\!
V_{\rm eff} &=&
V(\phi^*\phi) +
V_2(m(\phi)) \nonumber\\
&=&
V(\phi^*\phi) +
 2\int_{|{\bf p}|\le P_F} \frac{d^3{\bf p}}{(2\pi)^3} \left[\omega(|{\bf p}|)-\mu\right],
\label{effp1}
\end{eqnarray}
where $P_F$ is  the Fermi momentum and  $\mu={\mathcal E}_F=\sqrt{P_F^2+m^2(\phi)}$
the associated Fermi energy. The fermion density is
\begin{equation}
n%2\int \frac{d^3{\bf p}}{(2\pi)^3} \frac{1}{1+e^{-\beta(\omega({\bf p})-\mu)}}
=2\int_{|{\bf p}|\le P_F} \frac{d^3{\bf p}}{(2\pi)^3}
=\frac{1}{3\pi^2}P_F^3\,.
\label{density}
\end{equation}
The momentum integral of Eq.~({\ref{effp1}})
yields with the abbreviations
$\hat m\equiv  m(\phi)/P_F$ and $V_2(m(\phi))\equiv (P_F^4/24\pi^2) v_2(\hat m)$ the result:
\begin{eqnarray}
v_2(\hat m)&\!=\!&
\Big\{\!-8\!+\!3 \sqrt{1+\hat m^2}(2+\hat m^2)\nonumber\\
&\!+\!&3\hat m^4\left\{
\log \hat m-\log(1\!+\! \sqrt{1+\hat m^2})\right]
\!\Big\}\,.
\label{@TAY0}
\end{eqnarray}
The reduced potential $v_2$
has the Taylor
expansion
\begin{equation}
v_2(\hat m)=
-2+6 \hat m^2 +\frac{3}{4}\hat m^4\left(1+2
\log
\frac{\hat m^2}4\right)
%-\frac{3}{4}\hat m^6
+\dots\, .
\label{@TAY}
\end{equation}
We shall 
ignore 
the bare masses $m_0$ of fermions in Eq.~(\ref{mphi}),
which makes $\hat m^2$ proportional to
$\phi^2$. From the quadratic term $6\,\hat m^2$ in Eq.~(\ref{@TAY}), we can immediately see that
for sufficiently
high $P_F$, the location of
the minimum of the combined field potential
(\ref{effp1}) will 
moves from the $\phi\not=0$-value
 of the broken U(1)-symmetry to the origin $\phi=0$,
thereby restoring
the spontaneously broken U(1)-symmetry.
This happens at
\begin{equation}
-\mu^2=
\frac{P_F^2}{4\pi^2}g^2\,,
\label{TRAN}\end{equation}
which can be estimated
by the quasicritical point where the curvature term changes
its sign.

If it were not for the $\hat m^4\log \hat m^2$-term in
Eq.~(\ref{@TAY}), the Taylor
expansion (\ref{@TAY}) would be of the Landau-type and the phase transition would be
of the second order.
The logarithmic term
destroys the Landau expansion.
Indeed, suppose $P_F$ is chosen so large that the
$\phi^2$ in Eq.~(\ref{effp1})
vanishes, at which the coefficient of $\phi^2$ in Eq.~(\ref{effp1}) changes from negative values to positive values. 
Then the fourth-order term 
\begin{eqnarray}
 \lambda \phi^4+\frac{1}{32\pi^2}g^4\phi^4 \left(1+2\log \frac{g^2\phi^2}4
\right)
\label{4-phi}\end{eqnarray}
has a minimum at
\begin{eqnarray}
\phi^2_s=\frac{4}{g^2}e^{-1/2}e^{-4/g^2}e^{- 18\pi^2\lambda /g^4}\,,
\label{newv}\end{eqnarray}
and the minimal value of the field potential is
\begin{eqnarray}
-\frac{1}{8}\phi_s^4\left(\lambda+\frac{2g^2}{\pi^2}\right)\,.
\label{newm}
\end{eqnarray}
For the fourth-order term coupling $\lambda\sim {\mathcal O}(1)$ and the Yukawa coupling $g\sim {\mathcal O}(1)$, $\phi_s\sim 0$ in unit of $P_F$,
this implies
that for slightly positive curvature,
the field {\em jumps\/}
from its nonzero expectation value
to zero
in a first-order phase transition. %(should be second order??) 
Note that the argument can only be used qualitatively,
as been emphasized
by Colemen and Weinberg \cite{coleman,weinberg}.
The reason is that the 
new minimum lies in a field regime
where the perturbation theory is no longer reliable \cite{HKN}.
%In that regime the effective potential has to be calculated in another way
%by a menthod that has only recently been developed \cite{HKN}.
%\red{The transition is only weakly of first order. why is the first order? it s%hould be the second order??}

The condition (\ref{TRAN}) yields the critical fermion-number density
\begin{eqnarray}
n_c=\frac{8\pi}{3} \left(\frac{\sqrt{-\mu^2}}{g}\right)^3=\frac{8\pi}{3} \left(\frac{\sqrt{-\mu^2} }{m}\right)^3\langle\phi\rangle^3
\,,
\label{@TRAN2}
\end{eqnarray}
where we express the Yukawa coupling $g$ in terms of the fermion
mass $m(\phi)$ and the expectation value $\langle\phi\rangle$ as
$
g=m\sqrt{- 2 \lambda /\mu^2}\,.
$
When the fermion number-density is smaller than this critical density $n_c$, we are in the broken phase of spontaneous symmetry breaking by the Nambu-Higgs-Englert mechanism. When the fermion number-density is larger than this critical density $n_c$, symmetries are restored and we are in the symmetric phase.  In our calculations, we do not consider the temperature and external magnetic field, however their effects are clear that the critical density decreases as the temperature increases, instead, as the magnetic field strength decreases.

We shall now apply these considerations
to the possible realizations in our Universe: (i) the standard model of particle physics in the early Universe; 
(ii) the Gell-Mann-L\`evy $\sigma$-model of nuclear physics and its 
application in the stellar dynamics; (iii) the superconductivity of 
condensate matter physics. 

%\vskip0.1cm
\noindent
{\bf The standard model for particle physics.}
\hskip0.1cm
The standard model of electroweak
interactions
 has the local gauge
symmetry $SU_L(2)\otimes U_Y(1)$ with respect to the fermions
of each generation, the lightest being
 ($e,\mu,u,d$).
The gauge fields $A_\mu^i$ $(i=1,2,3)$ of the
left-handed
gauge group $SU_L(2)$ with the coupling $g$,
and the gauge field $B_\mu$ of the hyper-charge $U_Y(1)$
with the coupling $g'$
form the combinations to
yield physical
$W^\pm$-, $Z$-bosons and
photon fields.
\comment{The electric charge $e$ is related to the gauge couplings
$g$ and $g'$ by $e =g'\cos\theta_W = g \sin\theta_W$, where the Weinberg angle
$\theta_W\approx 28.74^0$ with $\sin^2\theta_W\approx 0.231$
Without spontaneous symmetry breaking,
the vector bosons and fermions masses are all zero.}

One introduces a doublet of complex scalar Higgs fields:
$
\phi = (\phi^+, \phi_0),
$
of hypercharge $Y=+1$
which is assumed to have the field potential
(\ref{hp}). For $\mu^2<0$,
 $\phi$ acquires nonvanishing vacuum expectation value, which may be assumed
to be real and to point in the direction
of
 $\phi_0$:
\begin{eqnarray}
\langle \phi\rangle = \frac{1}{\sqrt{2}}\left(\matrix{0\cr v}\right)+
{\mathcal O}(\hbar),~ v= \sqrt{2}\langle \phi\rangle=\sqrt{-\mu^2/ \lambda }\,\,.
\label{higg2}
\end{eqnarray}
The nonabelian
symmetry is broken, but the symmetry under the abelian subgroup
$U_{\rm em}(1)$ is
preserved. The $W^\pm$-, $Z^0$-vector bosons and fermions acquire their masses.
The size of $v\approx246.2\,{\rm GeV}\,$ is fixed by the measured masses of the vector bosons $v=2M_W\sin \theta_W/e$, where $\cos\theta_W =M_W/M_Z$, 
\comment{Most directly, we
may use the Fermi coupling constant
\begin{equation}
{G_F}=\frac{ \sqrt{2} g^2}{8M_W^2}=\frac{1}{ \sqrt{2} v^2}=1.16637(1)\times
10^{-5}{\rm GeV}^{-2}
\label{Fermig}\end{equation}
to find $v$:
\begin{equation}
v\approx246.2\,{\rm GeV}\,.
\label{evev}\end{equation}}
The size of the Higgs mass $m_{_H}$ determines $ -\mu^2 $
to be
\begin{equation}
-\mu^2=m_{_H}^2/2, \quad m_{_H}^2/2=\lambda v^2\,.
\label{higgsm}
\end{equation}
The phase transition takes roughly place
if the condition (\ref{@TRAN2}) is fulfilled:
\begin{equation}
n=n_c=
\frac{8\pi}{3} \left(\frac{\sqrt{-\mu^2}}{2m\lambda}\right)^3
=\frac{8\pi}{3} \left(\frac{m_{_H}}{2m}\right)^3v^3\,.
\label{@TRAN3}
\end{equation}
Suppose that only top-quark mass $m_t$ is originated from the spontaneous symmetry breaking, while other fermions masses from the explicit symmetry breaking \cite{xue2013}, then the critical density (\ref{@TRAN3}) yields,
\begin{equation}
n_c=
\frac{8\pi}{3} \left(\frac{m_{_H}}{2m_t}\right)^3v^3\approx   5.89\times 10^{6}{\rm GeV}^3\,.
\label{@TRANt}
\end{equation}
where $m_t\approx 173.1$GeV and $m_{_H}\approx 126\,$GeV.

The number-density that is larger than the critical density 
%(\ref{@TRAN3v}) or 
(\ref{@TRANt}) can only be realized in the early Universe, which starts from the Planck density $M_{\rm planck}^3\sim 10^{57}{\rm GeV}^3$, and the critical density 
%(\ref{@TRAN3v}) or 
(\ref{@TRANt}) is in the era of electroweak interactions of the temperature $10^2-10^7$GeV.   All gauge symmetries are preserved, all fields of gauge bosons, fermions and Higgs are massless, they are radiation fields. There is no any mass-threshold for pairs of particle-antiparticle creation, and the pair-production rate is greatly enhanced. All energy is deposited in radiation fields of massless particles. Beside, because of massless $W^\pm$- and $Z^0$-bosons, the cross-sections of weak-interaction become ($g=e/\sin\theta_W$)
\begin{eqnarray}
\sigma_{\rm weak} \approx \frac{4\alpha^2}{\sin^4\theta_W\,E^2}\,,
\label{cross2}
\end{eqnarray}
which is larger than the cross-sections $\sigma \approx \alpha^2/E^2$ of 
electromagnetic interaction in high energies $E$, where $\sin^2\theta_W$ 
is no longer $0.231$ but smaller than one. This could bring
an immense enhancement of the reaction speeds 
of the elementary processes. 
As one of consequences, all neutrinos are thus trapped inside the
system. This implies that in this symmetric phase of high density, all fields are radiation fields giving very high pressure on the system for then expansion. As the fermion number-density decreases, the system undergoes the phase transition to the broken phase where fermions and $W^\pm$- and $Z^0$-bosons become massive. This phase transition should have dramatic effects on the early Universe evolution.
These discussions can be generalized to the spontaneous breaking scales of the theory of the $SU(5)$ (or $SO(10)$) grand unification or supersymmetry.

%\vskip0.1cm
\noindent
{\bf The Gell-Mann-L\'evy-model for nuclear physics.}
\hskip0.1cm
The forces between nucleons 
can very well be described by an effective Lagrangian 
involving the  nucleon fields proton $p$ and neutron $n$
coupled to various meson fields. The main effects of this model 
become visible by concentrating upon the nuclear and meson fields alone,
which historically was done by Gell-Mann and L\`evy 
\cite{GML} in their
famous $\sigma$-model in 1960. It contains a fermion doublet $\psi=(p,n)$ 
of proton and neutron, whose bare masses are $m_{_N}^0$, a triplet of pseudo-scalar pions ${\vec \pi}$, and a scalar field $\sigma$. The corresponding Lagrangian is written as
\begin{eqnarray}
{\mathcal L}%(\bar\psi, \psi,\sigma,\vec \pi)
&=& \bar\psi[i\gamma_\mu\partial^\mu-m_{_N}^0-g(\sigma +i\vec \pi \cdot \vec \tau \gamma_5)]\psi 
\nonumber\\
&+&\frac{1}{2}[(\partial\vec \pi)^2+(\partial\sigma)^2] - V(\sigma,\vec \pi)\,,\label{sigma}
%\\V(\sigma,\vec \pi)&=&\frac{\mu^2}{2}(\sigma^2 +\vec \pi^2)
%+\frac{\lambda}{4}(\sigma^2 +\vec \pi^2)^2,
\end{eqnarray}
with 
$V(\sigma,\vec \pi)=\frac{\mu^2}{2}(\sigma^2 +\vec \pi^2)
+\frac{\lambda}{4}(\sigma^2 +\vec \pi^2)^2$.
This exhibits an $SU(2)$-symmetry to accommodate baryons $(p,n)$ in 
its fundamental representations and 
mesons $(\sigma, \vec \pi)$ in its adjoint representation O(3).
Analogous calculations for this Lagrangian lead to the effective potential 
$V_{\rm eff}(\sigma,\vec \pi)$ in the form of Eq.~(\ref{effp}) with the 
replacement $m^2(\phi)\rightarrow (m_{_N}^0+g\sigma)^2+g^2\vec \pi^2$.

When the nucleon density is small and its effect on the field potential is 
negligible, the $\sigma$-model undergoes the spontaneous symmetry breaking 
for $\mu<0$, developing expectational values $\langle \sigma\rangle =v/
\sqrt{2}, \, v=\sqrt{-\mu^2/\lambda}$ and $\langle \vec \pi\rangle=0$. The 
particle spectra are the massive $\sigma$-field $m^2_\sigma=-2\,\mu^2$, 
three Goldstone pions $m_\pi=0$, and nucleons of physical masses $m_{_N}=m_
{_N}^0+g\langle \sigma\rangle$. Recall that applying this model to the 
up and down quarks, the bare quark masses
$m_0$ are of the order of 10\,MeV, while the physical 
quark masses $m$ is of the order of 1/3 of the 
nucleon mass. The model was successful 
in explaining the
smallness of the pion mass, PCAC,  the pion decay,
as well as the low-energy theorems for pion-pion or
the pion-nucleon scattering. 

As the density of nucleons increases,  the field potential $V(\sigma,\vec \pi)$ is changed to $V_{\rm eff}(\sigma,\vec \pi)$ of Eq.~(\ref{effp}), the phenomenon of phase transition discussed above must occur at the critical density near to the nuclear density
\begin{equation}
n_c
=\frac{\pi}{3} \left(\frac{m_\sigma}{m_{_N}-m_{_N}^0}\right)^3v^3\approx \frac{\pi}{3} \left(\frac{m_\sigma}{m_{_N}}v\right)^3\approx n_{\rm nucl},
\label{@TRANn}
\end{equation}
where we consider $m_{_N}\gg m_{_N}^0$, $m_\sigma\approx 400$MeV and $v\approx 283\,$MeV for $\lambda\sim {\mathcal O}(1)$. In the symmetric 
phase, the expectational value $\langle \sigma\rangle=0$ or the small 
value of Eq.~(\ref{newv}), from the quadratic terms of nucleon and meson 
fields in the Lagrangian with effective potential 
$V_{\rm eff}(\sigma,\vec \pi)$, we obtain that 
nucleons have their bare masses $m_{_N}^0$, and the 
$\sigma$- and $\vec \pi$-fields have their effective 
masses $m^2_{\sigma,\pi} = \mu^2 + (gP_F/2\pi)^2 >0$. The $SU(2)$-symmetry is completely restored, however, the symmetry is realized by different massive spectra of baryons and mesons.
This second-order phase transition forms a sharp surface of a nucleus 
with the width $\sim m^{-1}_{\sigma,\pi}$, 
that distinguishes the broken-symmetry phase, where nucleons and mesons interact as a dilute gas, from the symmetric phase, where nucleons and mesons couple together as a liquid at the nuclear density. In 
the formation or collapse of stellar cores near to the nuclear density, 
as the nucleon density increases due to an 
attractive gravitational force and becomes overcritical, %(\ref{@TRANn}), 
the second-order phase transition may take place, the massive spectra of 
particles are changed and baryons may become relativistic particles 
if $m_{_N}^0\ll P_F$, which may result in important 
physical consequences. In addition, 
since electrons do not associate to this transition, strong electric 
fields over the critical value $E_c=m^2_e/e^2$ of Sauter, Euler and 
Heisenberg could possibly be developed \cite{rx2013}.

%\vskip0.1cm
\noindent
{\bf The superconductivity of condensed matter.}
\hskip0.1cm
To study the effects of fermion density on the 
spontaneous symmetry breaking and Cooper-pair formation in 
the BCS model for superconductivity, the Euclidean action (\ref{l1}) should be 
\begin{eqnarray}
A\!&=&\!\int_0^\beta d\tau\int d{\bf x} \Big[{\mathcal K}\!-\! g\bar\psi_\uparrow(x) \bar\psi_\downarrow (x)\psi_\downarrow (x)\psi_\uparrow(x) \Big],
\label{sca}
\end{eqnarray}
where $x=({\bf x},\tau)$, the kinetic term 
${\mathcal K}=\bar \psi_\sigma (x)[\partial_\tau -\nabla^2/2m -\mu]\psi_\sigma(x)$ and 
the attractive interaction $g$. The standard approach of introducing $\Delta(x)=\psi_\downarrow (x)\psi_\uparrow(x) $, and expressing 
Eq.~(\ref{sca}) in terms of quadratic fermion fields,   integrating out fermion fields leads the effective action (see for example Refs.~\cite{MRE1993,ERM1997,kleinertsc,KLPP}) 
\begin{equation}
A_{\rm eff}\!=\!\int_0^\beta d\tau\int d{\bf x} 
\left\{\frac{|\Delta(x)|^2}{g}\! -\! \frac{1}{\beta}{\rm Tr}\,{\rm ln}\,(\beta {\bf G}^{-1}) \right\},
\label{esca}
\end{equation}
where ${\bf G}^{-1}[\Delta(x)]$ is the inverse operator of quadratic fermion fields. Adopt the approximation of 
a uniform static saddle point $\Delta(x)\approx \Delta_0$, 
that satisfies the saddle-point condition
$\delta A_{\rm eff}(\Delta_0)/\delta \Delta_0=0$, which can be written as the 
renormalized gap-equation
\begin{equation}
\frac{m}{4\pi a_s}\!=\!\sum_{\bf k} \left[\frac{1}{2\epsilon_{\bf k}}\!-\!\frac{\tanh (\beta E_{\bf k}/2)}{2E_{\bf k}}\right],~ \frac{m}{4\pi a_s}=\sum_{\bf k}
\frac{1}{2\epsilon_{\bf k}}-\frac{1}{g} ,
\label{gapsc}
\end{equation}
where $E_{\bf k}
=\sqrt{\xi_{\bf k}^2+\Delta_0^2}$, $\xi_{\bf k}=\epsilon_{\bf k}-\mu$ and $\epsilon_{\bf k}=|{\bf k}|^2/2m$. The second equation in Eq.~(\ref{gapsc}) regulates the ultraviolet behaviors in the gap-equation by 
going from $g$ to the renormalized coupling $g_R$ 
in terms of the observable s-wave scattering length $a_s$.
The fermion number $N=-\partial \Omega/\partial \mu\approx -\partial \Omega(\Delta_0)/\partial \mu$ leading to the number equation 
\begin{equation}
n=\frac{1}{V}\sum_{\bf k}\left[1- \frac{\xi_{\bf k}}{E_{\bf k}}\tanh \left(\frac{\beta E_{\bf k}}{2}\right)\right].
\label{nsc}
\end{equation}
For the sake of simplicity to illustrate the critical density, we consider the zero-temperature case $T=0$, 
$\mu=\E_F=P_F^2/2m=(3\pi^2n)^{2/3}/2m$. 
In the weak coupling BCS limit $1/(k_Fa_s)\rightarrow -\infty$, 
solving Eqs.~(\ref{gapsc}) and (\ref{nsc}) one obtains \cite{ERM1997} that 
$\Delta_0(0)/\E_F = 1.1 \exp [-\pi/(2P_F|a_s|)]$.
This implies that $\Delta_0(0)/\E_F\approx 0$ for $P_F\approx (2a_s)^{-1}$, i.e., the critical density for restoring the symmetry
\begin{equation}
n_c \approx \frac{1}{3\pi^2}\left(\frac{1}{2a_s}\right)^3.
\label{crisp}
\end{equation}
This physically means that a restoration of the spontaneously 
broken symmetry may be caused by the degenerate pressure of fermions 
[see Eq.~(\ref{fp})],
with the dramatic consequence of destroying the supercurrent. Although 
calculations are made in the zero-temperature case,
we can expect
 that the presence of temperature will enhance restoration,
since above $T_c$, 
superconductivity is completely destroyed.

%\vskip0.1cm
\noindent
{\bf Conclusion and remarks.}
\hskip0.1cm
In this Letter we point out the fact that the number density or 
degenerate pressure of many-fermion systems plays the same role of 
temperature acting on a spontaneous breakdown of symmetries, 
and the increase of fermion density implies the decrease of the value of 
critical temperature $T_c$ for a phase transition. We obtain the critical 
fermion density in the zero-temperature case   
and discuss possible observable consequences in our Universe.  

To end this Letter, we draw attention to the fact that
the phenomena act in precisely the opposite directions of what we see in many-boson systems upon a spontaneous breakdown of a symmetry.
As far as the ground state is concerned, this has been well known
since the beginning of quantum field theory.
A boson field at zero temperature in the vacuum 
has zero-point energy $\omega \hbar/2$,
where $\omega=\sqrt{{\bf p}^2+m^2}$, whereas fermions have the same energy with the opposite sign.
But that the effects are opposite also 
in interacting systems deserves to be emphasized.
What has been known since its discovery,
the famous phenomenon of Bose-Einstein Condensation 
(BEC) is caused by squeezing at low temperature 
too many bosons into a small volume.
Thus the exceeding of a critical density leads to a nonzero expectation value of the boson field, i.e.,
to a spontaneous breakdown of the symmetry of particle number conservation. 
This is precisely the opposite direction of the 
influence of a 
fermionic particle density, which might be seen from the opposite sign 
of the fermionic pressure (\ref{fp}) and the bosonic pressure (\ref{bp}). 
It is interesting to see these phenomena in the quark-gluon plasma due to opposite effects of quark density and gluon density.

\end{document}